\documentclass{jpp}
\usepackage{url}
\usepackage{graphicx}
\usepackage{subfig}
\usepackage{appendix}
\usepackage{epstopdf}
\usepackage{natbib}
\usepackage{tikz}
\usepackage[all]{xy}

\begin{document}

\title{Generalised quasi-linear approximation of the HMRI}
\author{Adam Child\aff{1}
  \corresp{\email{mm08ac@leeds.ac.uk}},
  Rainer Hollerbach\aff{1},
  Brad Marston\aff{2}
 \and Steven Tobias\aff{1}}

\affiliation{\aff{1}Department of Mathematics, University of Leeds, Leeds, LS2 9JT, UK
\aff{2}Department of Physics, Box 1843, Brown University, Providence, RI 02912, USA}

\maketitle

\begin{abstract}
Motivated by recent advances in Direct Statistical Simulation (DSS) of astrophysical phenomena such as out of equilibrium jets, we perform a Direct Numerical Simulation (DNS) of the helical magnetorotational instability (HMRI) under the generalised quasilinear approximation (GQL). This approximation generalises the quasilinear approximation (QL) to include the self-consistent interaction of large-scale modes, interpolating between fully nonlinear DNS and QL DNS whilst still remaining formally linear in the small scales. In this paper we address whether GQL can more accurately describe low-order statistics of axisymmetric HMRI when compared with QL by performing DNS under various degrees of GQL approximation. We utilise various diagnostics, such as energy spectra in addition to  first and second cumulants, for calculations performed for a range of  Reynolds and Hartmann numbers (describing rotation and imposed magnetic field strength respectively). We find that GQL performs significantly better than QL in describing the statistics of the HMRI even when relatively few large-scale modes are kept in the formalism. We conclude that DSS based on GQL (GCE2) will be significantly more accurate than that based on QL (CE2). 

\end{abstract}

\section{Introduction}

The magnetorotational instability (MRI) \citep{Velikhov} is one of the most important processes in modern astrophysics, thought to be responsible for the outward transport of angular momentum in magnetised accretion disks \citep{balbus1991}, and has been the subject of numerous publications in recent years (see the review by \citet{julien2010}). It is driven by the interaction of large scale Keplerian rotation and mean magnetic fields, and the nonlinear saturation of the instability can lead to turbulent dynamics and efficient angular momentum transport. The presence of an axial magnetic field enables the instability to proceed for fluids with angular momentum profiles that increase with radius; for the same configurations hydrodynamic flows are stable via the Rayleigh criterion.

Direct numerical simulations of this instability \citep{Suzuki2014,Bai2014,Meheut2015,Gressel2015}, though illuminating for the underlying physics, are performed at parameter values that are far from those pertaining to astrophysical disks. For this reason, approaches other than direct solution of the equations have been brought to bear on this problem. The first of these is the utilisation of laboratory experiments in parameter regimes that may not be accessible to numerical simulations. This approach has revealed much about the interaction of rotating fluids, turbulence and magnetic fields as described below, but its direct applicability to the problem of angular momentum transport in disks remains problematic. The second strategy is to utilise Direct Statistical Simulation (DSS) for astrophysical flows \citep{tobias2011}, where the statistics of such flows are obtained directly rather than emerging from averaging of the dynamics from DNS. 

As noted above, an important tool to understand such an instability is laboratory experiments. The first of these, by \citet{rudiger2001} and \citet{ji2001} simultaneously but independently, proposed an experimental set-up with  a liquid metal in a cylindrical Taylor-Couette geometry with an imposed axial magnetic field. Instability occurs when the non-dimensional magnetic Reynolds number $\mathrm{Rm}=\Omega_i r_i^2/\eta\ge O(10)$. Together with the fact that liquid metals have magnetic Prandtl number $\mathrm{Pm}\sim O(10^{-6})$, this translates into a hydrodynamic Reynolds number $\mathrm{Re}=\mathrm{Rm}/\mathrm{Pm}\ge O({10^7})$.  Such large values present severe technical difficulties \citep{Schartman2009}, especially when it comes to conditions at the end-plates, which have a strong tendency to dominate the flow \citep{hollerbach2004,avila2012,gissinger2012}. As a result, this `classical' set-up has not yet succeeded in obtaining the standard MRI (SMRI) -- although the Princeton MRI experiment has yielded many other interesting results \citep{nornberg2010,roach2012}. Very recently it has also been suggested \citep{Flanagan2015} that the SMRI could be obtained using plasma rather than liquid metal.

Alternatively, \citet{hollerbach2005} suggested adding an azimuthal field $B_\phi\sim r^{-1}$, imposed by an electric current running along the central axis of the cylinders. This has a surprisingly large influence on the onset of instability, switching the criterion from $\mathrm{Rm}\ge O(10)$ to $\mathrm{Re}\ge O(10^3)$ -- that is, reducing the critical Reynolds number by some four orders of magnitude.  Despite this dramatic reduction, and indeed fundamentally different scaling with $\mathrm{Pm}$, this new instability, now known as the helical MRI (HMRI), is continuously connected to the SMRI (although a number of subtleties are involved \citep{kirillov2010,kirillov2014}). Following its theoretical prediction, the HMRI was quickly observed in a series of experiments in the PROMISE facility (see \citep{stefani2009} and earlier references therein). The HMRI also continues to be the subject of further theoretical analysis \citep{Liu2006,Priede2007,Priede2011}.

The main purpose of this paper, however, is to determine the regime of applicability of the GQL approximation for a system of wall-bounded, electrically conducting flows. The HMRI is an excellent system to elucidate the strengths and weaknesses of this approach as it shows turbulent dynamics even in its axisymmetric configuration described above.  The GQL representation has been examined in the context of two-dimensional driven turbulence on a spherical surface and $\beta$-plane and shown to be more effective than the regular QL approximation in reproducing both the dynamics and statistics of these flows away from equilibrium \citep{mct16}. The importance of establishing the regime of applicability of various approximations arises owing to their use in deriving statistical theories.

As noted by \citet{tobias2011}, the enormous range of spatial and temporal scales present in many astrophysical phenomena poses a problem for DNS in that each scale must be properly resolved; this is unlikely, even on today's powerful parallel computers --- and indeed those of the foreseeable future. The problem is that in most astrophysical applications, the large-scale dynamics are influenced by, and in turn influence, the small-scale interactions. 

A complementary approach, that has been afforded much attention recently is that of Direct Statistical Simulation (DSS), in which the low-order statistics are obtained directly from a hierarchy of cumulant equations, and a closure is applied to this hierarchy in order to provide tractable calculations. DSS has a number of advantages over DNS, chiefly that the statistics vary slowly in time and are smoother in space and can be explored via the evolution of relatively few modes on a slow manifold.
There have been a number of examples of Direct Statistical Simulation in recent works, most notably involving various planetary jets; \citet{tobias2011} derived a formulation for flow on rotating spherical surface and compared with the DNS solutions showing considerable accuracy. Indeed, DSS is shown to be useful in analysing the underlying physics of flows in that it in this case demonstrates that jets may be formed without an inverse cascade.  Moreover, quasilinear statistical framework has allowed \citet{squire2015} to probe the MRI dynamo in a shearing box in ways inaccessible in conventional DNS. All of the above investigations used the most simple form of DSS (known as CE2) --- this approximation is the statistical representation of the quasilinear approximation (QL). At this quasilinear level of approximation DSS is formally equivalent to the Stochastic Structural Stability Theory (SSST or S3T) of Farrell, Ioannou and collaborators \citep{farrellioannou2007,farrellioa08,Constantinou:2013fh}, which has been extensively used to model a number of physical systems and can be justified for systems near equilibrium for which there is a separation of timescales \citep{Bouchet2013}.

Of course, this is not to imply that quasilinear DSS (i.e. CE2) is universally applicable; there are a number of examples of inconsistencies with fully nonlinear DNS (NL DNS), evident for example in out-of-equilibrium jets studied with DSS for a number of parameter combinations. Far from equilibrium, CE2 ceases to be an adequate representation of the system, as shown by \citet{tobias2013}. In that paper it was demonstrated that when the system was in statistical equilibrium and thus dominated by strong jets, CE2 does in fact do a good job of reproducing both the number and strength of jets in the system. However, the further away from equilibrium, the worse CE2 is seen to perform, resulting in both inaccurate jet strengths and inaccurate numbers of jets (see also \citet{srinivasan2012}). Two options have been suggested to remedy the situation. The first is to include eddy-eddy scattering in the truncation. This leads to the CE3 (or CE2.5) approximation, which has been shown to improve performance of DSS away from equilibrium \citep{mqt2016}. 

A second possibility is to generalise the quasilinear approximation \citep{mct16} and derive a  corresponding statistical theory (GCE2). We explore the accuracy of such direct statistical simulation of the HMRI at GCE2 by performing a series of generalised quasilinear DNS (QL DNS) in the inductionless limit. First we formulate the model and give details of the methods utilised for DNS. We continue by defining the GQL approximation and compare its efficacy with that of QL theory by examining how well it reproduces a number of diagnostic quantities, such as the energy spectra and first and second cumulants of the flow. We conclude by briefly discussing the implications of our results to possible statistical simulations.

\section{Formulation and Numerical Method}

We consider axisymmetric cylindrical Taylor-Couette flow, with radii and angular velocities at the inner and outer cylinders given by $r_i, r_o$ and  $\Omega_i$ and $\Omega_o$  respectively. We set $r_i/r_o=0.5$ to coincide with the aspect ratio of the PROMISE experiment. Unlike an experiment though, with its inevitable end-plates, we take the cylinders to be  axially periodic  with a basic periodicity of 40 times the gap-width. This is sufficiently long to allow turbulent structures to develop in the axial direction; it is the dynamics and interactions of these structures that we then wish to explore in this work.

The flow and magnetic field are described by the Navier-Stokes and induction equations,
\begin{eqnarray}
\mathrm{Re}\frac{\partial{{\bf U}}}{\partial{t}} - \nabla^{2}{\bf U} &=& -\nabla{p} - \mathrm{Re} {\bf U}\cdot\nabla{{\bf U}} + \mathrm{Ha}^{2}\mathrm{Rm}^{-1}(\nabla\times{{\bf B}})\times{{\bf B}},\\
\frac{\partial{{\bf B}}}{\partial{t}} &=&  \nabla\times({\bf U} \times{{\bf B}})+\mathrm{Rm}^{-1}\nabla^{2}{\bf B},
\end{eqnarray}
where we have non-dimensionalised as follows: we scale lengths with the gap-width $(r_o-r_i)\hat{r}$, time on the rotational time-scale $t=\Omega_i^{-1}\hat{t}$, $U$ proportional to length over time, $U=(r_o-r_i)\Omega_i\hat{U}$, and $B=B_z\hat{B}$, with $B_z$ the strength of the imposed axial magnetic field. The nondimensional quantities
\begin{equation}
 \mathrm{Rm}=(r_o-r_i)\Omega_i/\eta,\qquad \mathrm{Re}=(r_o-r_i)\Omega_i/\nu, \qquad \mathrm{Ha}=B_z(r_o-r_i)/\sqrt{\eta\nu\mu_0\rho}
\end{equation}
are the magnetic and hydrodynamic Reynolds numbers and the Hartmann number respectively.

As noted above, for the SMRI it is crucial that $\mathrm {Rm}$ be retained, and exceed $O(10)$. In contrast, the HMRI continues to exist even in the so-called inductionless limit $\mathrm{Rm}\to0$. Splitting the magnetic field into imposed and induced fields as ${\bf B}= {\bf B_0} + \mathrm{Rm}\,{\bf b}$ and then letting $\mathrm{Rm}\to0$ yields
\begin{eqnarray}
\mathrm{Re}\frac{\partial{{\bf U}}}{\partial{t}} - \nabla^{2}{\bf U} &=& -\nabla{p} - \mathrm{Re}{{\bf U}}\cdot\nabla{{\bf U}} + \mathrm{Ha}^{2}(\nabla\times{{\bf b}})\times{{\bf B_0}},\\
0 &=& \nabla^{2}{\bf b} + \nabla\times({\bf U} \times{{\bf B_0}}).
\end{eqnarray}
Note also that in this limit the only nonlinearity arises from the inertial term ${{\bf U}}\cdot\nabla{{\bf U}}$.

The imposed field is given by ${\bf B_0} ={\bf e}_z + \beta r^{-1}{\bf e}_\phi$; the HMRI operates when $\beta\ge O(1)$ \citep{hollerbach2005}. The rotation ratio is fixed at $\Omega_o/\Omega_i=0.27$, slightly beyond the Rayleigh value 0.25.  That is, for $\Omega_o/\Omega_i<0.25$ one would obtain purely hydrodynamic instabilities, which we are not interested in here, but for $\Omega_o/\Omega_i>0.25$ one obtains only the HMRI, for appropriate values of the three governing parameters $\mathrm{Re}$, $\mathrm{Ha}$, and $\beta$.

The boundary conditions on $\bf U$ are no-slip, as they must be for any Couette flow. For $\bf b$ the HMRI exists in much the same form for either insulating \citep{hollerbach2005} or perfectly conducting \citep{rudiger2006} boundaries. Since Fig.~2 of \citet{rudiger2006} mapped out a particularly thorough set of linear onset curves for perfectly conducting boundary conditions we will here consider only perfectly conducting boundaries.

Because the HMRI is an intrinsically axisymmetric phenomenon (just like the SMRI), we restrict attention to axisymmetric solutions, and further decompose as
\begin{equation}
{\bf U} = v \,{\bf e_{\phi}} + \nabla\times(\psi\,{{\bf e_\phi}}), \qquad {\bf b} = a \, {\bf e_{\phi}} + \nabla\times(b \,{\bf e_{\phi}}),
\end{equation}
thereby automatically satisfying $\nabla\cdot{\bf U}=\nabla\cdot{\bf b}=0$.
 
The full details of the numerical solution for $v$, $\psi$, $a$ and $b$ are presented in \citet{hollerbach2008}. Here we merely note that the radial structure is expanded in 50 Chebyshev polynomials, and the axial structure in 600 Fourier modes of the form $\exp(i\kappa z)$, where $\kappa=2\pi k/z_0$, with $z_0=40$ the imposed basic periodicity. The essence of the GQL again is to study interactions between different Fourier modes, so such a decomposition is convenient both numerically and theoretically.

\begin{figure}
 \centering
 \includegraphics[width=0.96\textwidth]{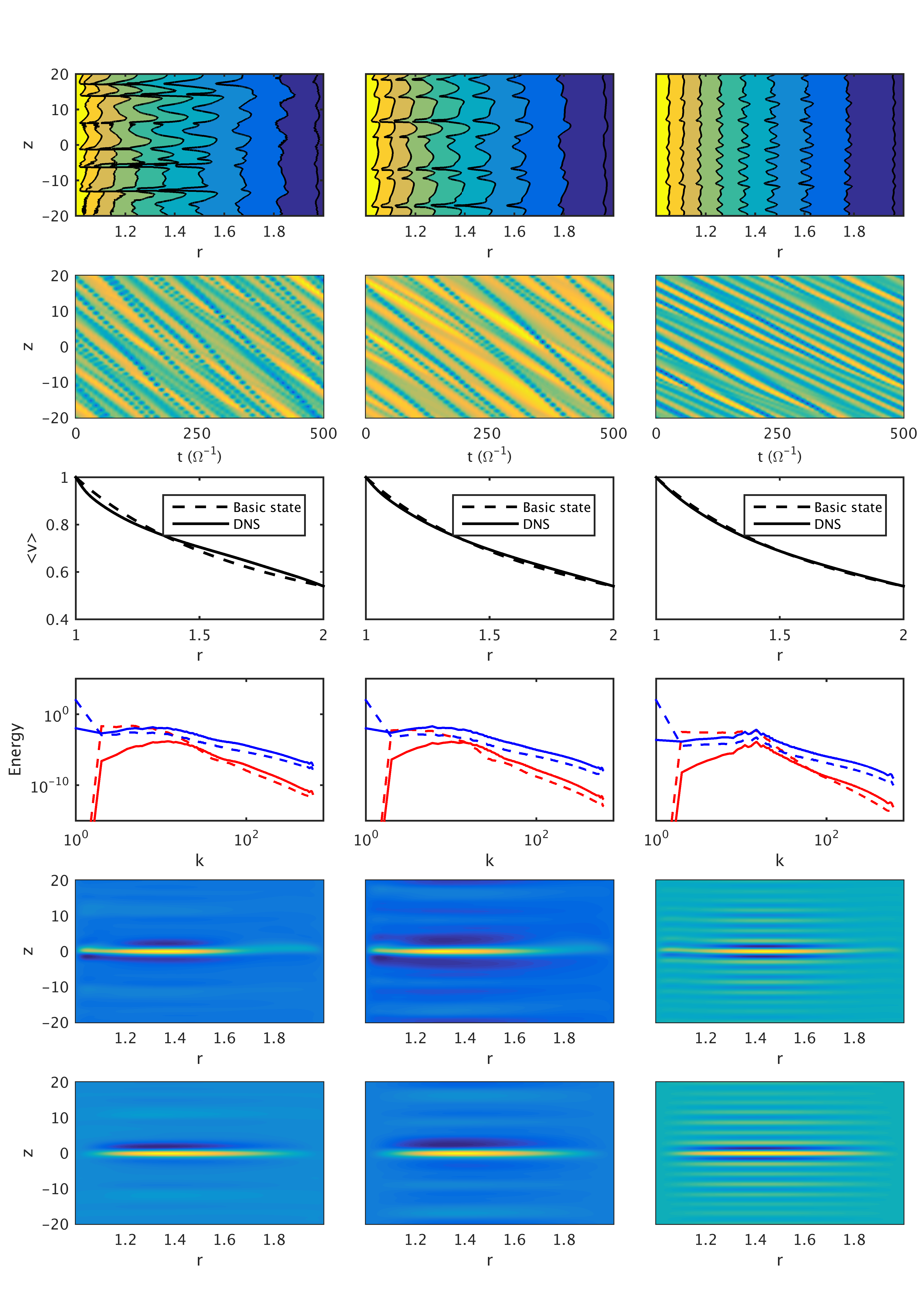} 
 \caption{Row: i) Contour snapshots of the azimuthal velocity in a radial cross section; ii) Hovm\"oller plots of $\psi$ as a function of $z$ and time; iii) Mean azimuthal flow (first cumulant) as a function of $r$;  iv) Vertical energy spectra for the magnetic field (red) and velocity (blue) with poloidal components (solid) and toroidal components (dashed); v) Second cumulants $\langle v' v' \rangle$;  vi) Second cumulants $\langle v' \psi' \rangle$ . Shown on the left is $\beta=5$, $\mathrm{Ha}=16$, in the centre is $\beta=2.5$, $\mathrm{Ha}=16$, and on the right is $\beta=10$, $\mathrm{Ha}=8$, each for $\mathrm{Re}=10^4$.}
\label{nldns}
\end{figure}

\subsection{Generalised quasilinear approximation}

In this section we outline how to apply the generalised quasilinear approximation introduced by \citet{mct16} to this axisymmetric HMRI system. For simplicity, let us consider the velocity ${\bf U}$, though similar considerations are applied to the perturbed magnetic field ${\bf b}$. In general the fields may be split via a standard Reynolds decomposition into mean and fluctuation (eddy) components, for example for the azimuthal velocity 
\begin{equation}
 v = {\overline{v}} + {v}'
\label{reynolds}
\end{equation}
where  overbar denotes a mean and prime denotes fluctuations about that mean. In previous papers the averaging procedure has involved taking a zonal mean, but in this axisymmetric geometry a natural definition of the mean is that part that is independent of the $z$-coordinate (i.e. the $\kappa=0$ mode). This standard definition for the mean and fluctuations (eddies) then naturally leads to the quasilinear approximation. In this approximation certain mode interactions are discarded from the dynamics. In particular whilst  mean/eddy $\rightarrow$ eddy and eddy/eddy $\rightarrow$ mean interactions are kept, eddy/eddy $\rightarrow$ eddy interactions are discarded. In this quasilinear approximation therefore only nonlocal (in wavenumber) elements of cascades or inverse cascades are possible \citep{tobias2011}

A similar decomposition is used for GQL. However here the velocity is  decomposed into large and small scale modes, where large and small scale are defined by the spectral wavenumber. That is we set
\begin{equation}
v(r,z)= \sum_{k=0}^{\Lambda}{v}_{lk}(r)e^{i\kappa z} + \sum_{k=\Lambda+1}^{N_k}{v}_{hk}(r)e^{i\kappa z} ,
\end{equation}
where the $v_l$ and $v_h$ are the large scale (denoted l for `low' modes) and smaller scale (denoted h for `high' modes)  respectively, and $\Lambda$ is the mode cut-off point between the two. It is clear then that when $\Lambda=0$ this expansion is the same as that for QL given in equation~(\ref{reynolds}). Hence the GQL formalism reverts to QL when $\Lambda=0$ as noted by \citet{mct16}

When $\Lambda \ne 0$ and the Generalised Quasilinear approximation differs from QL, the triad interactions that are retained and discarded are selected so as to satisfy the relevant conservation laws and enable closure \citep{mct16}. 
Briefly this involves retaining low/low $\rightarrow$ low, high/high $\rightarrow$ low and low/high $\rightarrow$ high interactions and discarding all others. Again retaining this set of interactions is consistent with QL when $\Lambda=0$. Furthermore as $\Lambda \rightarrow \infty$ the system consists solely of low modes and is formally the same as Direct Numerical Simulation. The aim of this paper is to determine how well GQL performs in reproducing the NL DNS results as $\Lambda$ is varied; clearly if it is able to do so for relatively small $\Lambda$ then this is extremely encouraging for producing Direct Statistical Simulation based on the GQL approximation.

\section{Results}

\subsection{Means of Comparison: Outputs of Interest}
As DSS is a statistical theory, we shall evaluate the GQL approximation by accumulating statistics using QL DNS and testing how well GQL reproduces the statistics of the HMRI (\citet{mct16} evaluated both the statistics and the dynamics for the problem of the driving of zonal jets). To that end we shall calculate average vertical spectra of the flow and magnetic field as well as the first and second cumulants for the flow.  The first cumulant is the mean over $z$ and time as a function of $r$ defined as
\begin{equation}
c_\psi(r)= \langle \psi(r,z,t)\rangle, \qquad c_v(r)= \langle v(r,z,t) \rangle, 
\end{equation}
where angled brackets indicate averaging over $z$ and $t$. the second cumulants yield information about the (non-local) fluctuation/fluctuation interactions and are defined as
\begin{eqnarray}
c_{\psi \psi}(r_1,r_2,z_1,z_2)&=& \langle \psi'(r_1,z_1,t)\psi'(r_2,z_2,t)\rangle,\\
c_{v v}(r_1,r_2,z_1,z_2)&=& \langle v'(r_1,z_1,t)v'(r_2,z_2,t)\rangle,\\
c_{v \psi}(r_1,r_2,z_1,z_2)&=& \langle v'(r_1,z_1,t)\psi'(r_2,z_2,t)\rangle.
\end{eqnarray}
The second cumulant can then be expressed as, for example
\begin{equation}
 c_{\psi \psi}(r_1,r_2,\xi) = \int \int \psi'(r_1,z_1)\psi'(r_2,z_1+\xi)\ dz_1 dt
\end{equation} 
where $\xi=z_2-z_1$. In this paper, we choose $r_1$ to be the point halfway between the two cylinders.

\subsection{Parameter Choice and Initialisation}

The generalised quasilinear approximation is straightforward to implement in the pseudo-spectral timestepping code -- the various triad interactions discussed by \citet{mct16} can be easily isolated in spectral space and discarded or retained as required.
We performed a number of simulations under various different parameter regimes for a number of truncations of the large scale modes $\Lambda$.

We initialise each simulation using the following procedure. For each parameter set we run a fully turbulent NL DNS solution until it reaches a statistically stationary state. It is this state that we use as initial conditions for runs at the same parameter values but with different cut-offs for the GQL approximation. We note that the same results are obtained if the GQL calculations are started from rest --- though these take longer to equilibrate. For each parameter set we evolve $\Lambda=0$ (corresponding directly to the QL expansion) and we also examine $\Lambda=1,2,3,6$ and $\Lambda=20$.  We evolve each of the equations using a timestep $dt = 5\times{10^{-3}}$, and run each simulation for a minimum of $8\times 10^4$ timesteps to make sure that the solutions reached are statistically steady. Any averages (over time) are then calculated using 300 equally spaced (in time) solutions, commencing from the 200000$^{th}$ step. 

We also give here some note about the rationale for our choice of model parameters ---  care has to be taken to make sure that Re is sufficiently above $\mathrm{Re}_c$, the critical Reynolds number. Intuitively, one might expect that the supercriticality of the system as measured by  $(\mathrm{Re} - \mathrm{Re}_c)$ will have bearing on the effectiveness of the QGL approximations, and so we choose to examine a number of different parameter sets. Based on Fig.~2 of \citep{rudiger2006}, we choose to focus on three specific choices; $\beta=5$, $\mathrm{Ha}=16$, $\beta=2.5$, $\mathrm{Ha}=16$ and $\beta=10$, $\mathrm{Ha}=8$, each for $\mathrm{Re}=5000,7000$ and $10000$. These choices ensure that the flow is sufficiently supercritical to generate a turbulent state. Of the three parameters sets, $\beta=5$, $\mathrm{Ha}=16$ has $\mathrm{Re}_c\approx 1000$, and is the most supercritical, and $\beta=2.5$, $\mathrm{Ha}=16$ has a slightly higher $\mathrm{Re}_c\approx1500$, and thus will give insight into how the GQL approximation holds for less turbulent flows. The point $\beta=10$, $\mathrm{Ha}=8$ is more interesting in that, even though $\mathrm{Re}_c<1000$, on the aforementioned figure the contours of $\mathrm{Re}_c$ are closely bunched together. That is, the flow would be viewed as being highly supercritical when the Reynolds parameter alone is examined, but close to marginal stability in terms of the magnetic field parameters. This is of particular interest in that it suggests that this parameter set may be only weakly nonlinear, for which we would expect even the standard QL approximation to give an accurate reproduction of the NL DNS. We claim that these three points and Re values sufficiently encapsulate the properties of the GQL approximation;  simulations for other parameters yield qualitatively similar results.

\subsection{NL DNS}

We begin by describing the behaviour of the system for NL DNS of the HMRI.
In figure 1, we show plots of various quantities -- contours of the azimuthal velocity, Hovm\"oller diagrams, mean azimuthal flows (first cumulants), energy spectra, and second cumulants -- for each parameter combination at the highest value of the Reynolds number $\mathrm{Re}=10000$. 
As seen in row 1 of figure 1, different parameter choices lead to varying degrees of turbulence, with $\beta=5$, $\mathrm{Ha}=16$ (first column)  being the strongest, and $\beta=10$, $\mathrm{Ha}=8$ (last column)-- that which we suspect is only weakly nonlinear -- being the weakest.

The HMRI is known to be a unidirectionally travelling wave at onset \citep{hollerbach2005,rudiger2006}, due to the $\pm z$ symmetry-breaking caused by an imposed field having both axial and azimuthal components \citep{knobloch}. This drifting nature is conveniently visualised by presenting a number of Hovm\"oller plots of $\psi(t,z)$ for a given radius, which allow the identification of the location of Taylor vortices. The streamfunction $\psi$ undergoes a reversal in sign when transitioning from one Taylor vortex to another. Therefore, simply plotting the sign of $\psi$ is equivalent to tracking the position of the vortices over time. It can be seen clearly that in DNS the vortices propagate downward in the axial direction as a travelling wave. Also apparent on these plots is the number of dislocations present in the flow; these are an effective measure of the strength of turbulence. It can be seen that in general, as expected, the number of dislocations decreases as $\mathrm{Re}_c$ increases for fixed Re, i.e. as the flow becomes less supercritical. However, at the point identified as only weakly nonlinear few of these nonlinear dislocations occur. The mean azimuthal flows shown in the third row demonstrate the change from the basic state. As expected the role of turbulence is to decrease the gradient of the mean azimuthal flow (removing the source of instability). For the more turbulent flows this change is more significant whilst for the less supercritical flows the departure from the mean state is much less pronounced as expected.

Turning our attention to the energy spectra, for which we plot the energy of both the toroidal and poloidal components of the magnetic field (in red) and velocity (in blue), we may note that regardless of the parameter regime, these are qualitatively very similar. The magnetic energy is smaller than that for the kinetic energy and has a steeper spectrum. In the weakly nonlinear case the energy peaks at a modenumber $k\approx 13$, which is similar to the linear mode of maximum growth-rate.  The energy spectra for $\mathrm{Re}=7000$ and $\mathrm{Re}=5000$ have similar characteristics to these plots. 
The second cumulants $\langle v' v' \rangle$ and $\langle v'  \psi' \rangle$ are very much localised in the axial direction, though there is strong radial correlation with the midpoint in radius. The axial localisation is a reflection of the flat energy spectra for the low modes, which leads to this $\delta$--function--like axial structure in the second cumulants. For the weaker turbulence (third column), the cumulants are different in that a periodic axial correlation is now apparent, such that flow at a point is correlated not only in the radial direction, but also periodically in each direction along the axis. A wavenumber of $k\approx 13$ in the second cumulants is observed, which corresponds to the peak observed in the energy spectrum.

These results will be used in the next section as benchmarks to evaluate the effectiveness of the GQL approximation at various truncations.

\subsection{GQL DNS}

\subsubsection{Weak turbulence: $\beta=10$ $\mathrm{Ha}=8$}

We start by examining the effectiveness of GQL DNS in comparison with QL DNS for the weak turbulence case $\beta=10$, $\mathrm{Ha}=8$, $\mathrm{Re}=10000$ (for which the DNS results are shown in the third column of Figure~1).
It is to be expected that the QL approximation will most likely have the greatest success in this parameter regime, as nonlinearities have smaller influence on the instability; recall QL is an exact representation of the dynamics in the formally weakly nonlinear regime.  We can see on comparing the Hovm\"oller plots (figure 2) that this expectation is correct; there are few dislocations in the NL DNS, meaning that even $\Lambda=0$ is an accurate approximation.

\begin{figure}
 \centering
 \includegraphics[width=\textwidth]{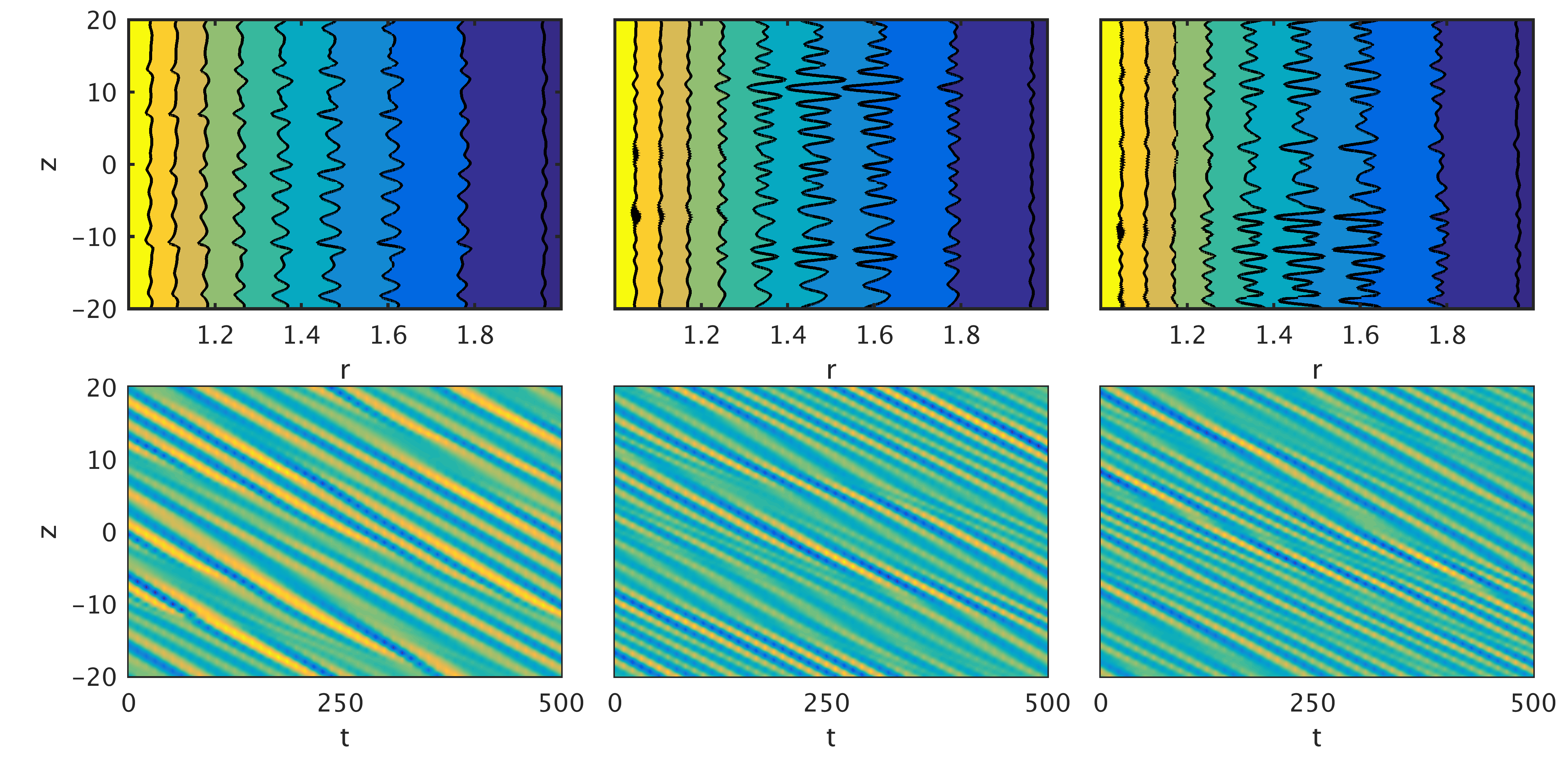}
 \caption{Snapshot contour plots of $u_\phi$ (top) and Hovm\"oller plots (bottom) for $\beta=10$, $\mathrm{Ha}=8$, $\mathrm{Re}=10^4$. Shown on the left is the NL DNS, in the middle is $\Lambda=0$, and on the right is $\Lambda=2$.}
\end{figure}

\begin{figure}
 \centering
 \includegraphics[width=\textwidth]{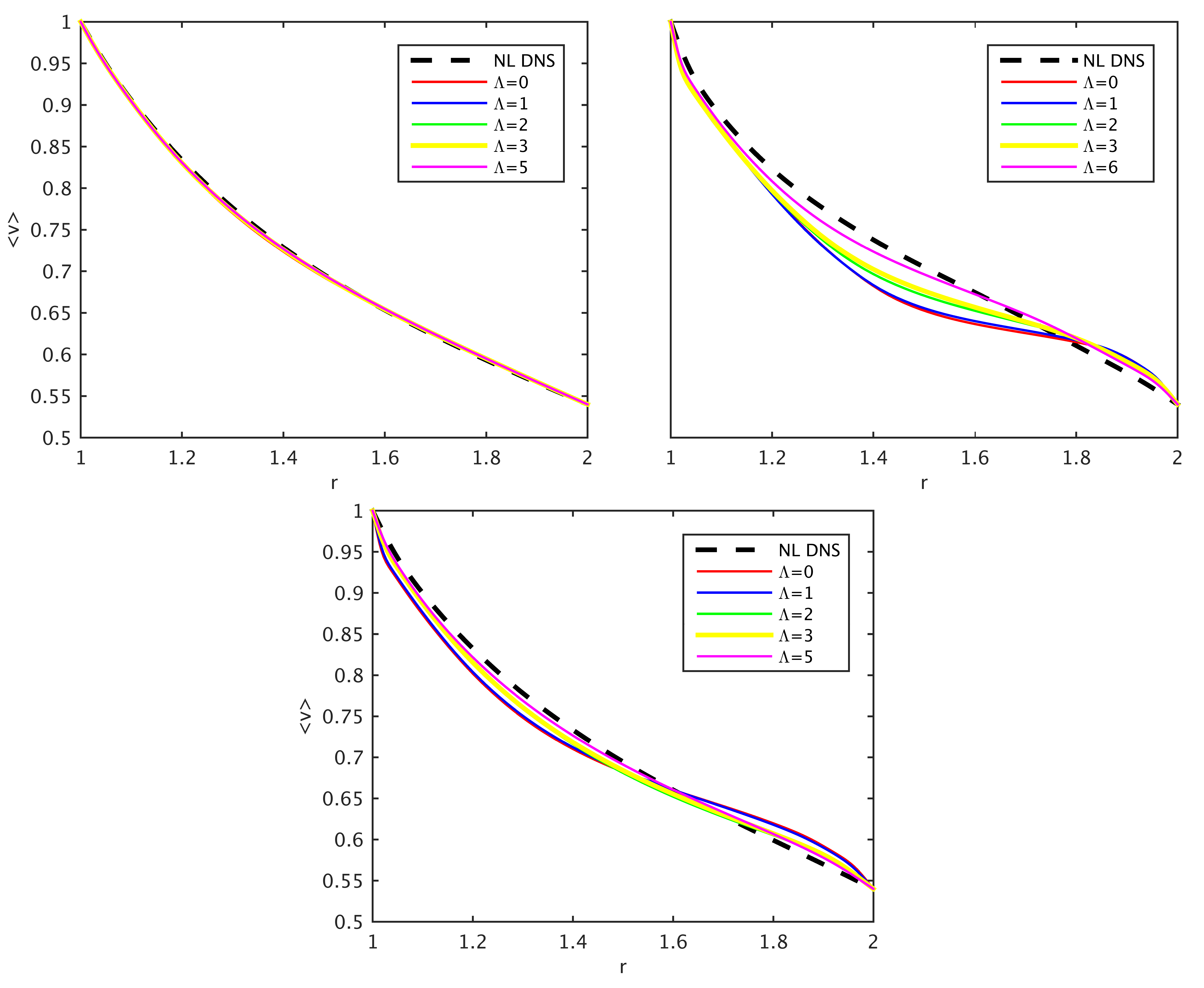}
 \caption{First cumulants $ \langle v \rangle$ for $\beta=10$ $\mathrm{Ha}=8$ (top left), $\beta=5$ $\mathrm{Ha}=16$ (top right) and $\beta=2.5$ $\mathrm{Ha}=16$ (bottom). Note that for $\beta=10$ $\mathrm{Ha}=16$, as the flow is only weakly nonlinear the NL DNS and GQL DNS profiles are near indistinguishable.}
\label{fcweak}
\label{fcstrong}
\label{fcmt}
\end{figure}

We can see from the first cumulants (figure~\ref{fcweak}(a)) that the various $\Lambda$ approximations are near indistinguishable from the NL DNS profile. Even the quasilinear $\Lambda=0$ gives an excellent approximation to the full solution. Indeed, for the first cumulants there is little advantage in using a GQL approximation over a QL approximation. Again for such a laminar solution this is perhaps to be expected.

\begin{figure}
 \centering
 \includegraphics[width=0.7\textwidth]{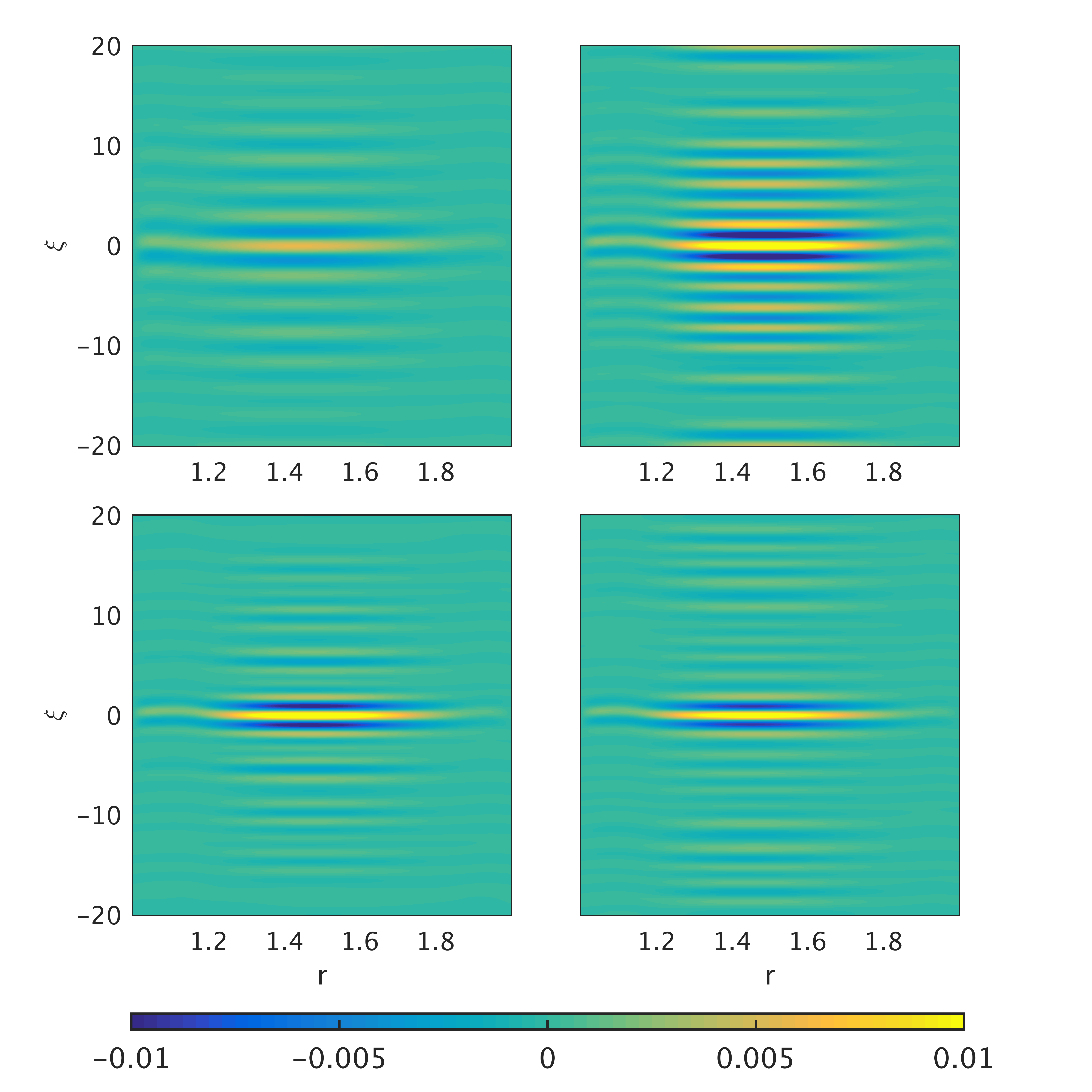}
 \caption{Second cumulants $\langle v' v' \rangle $ at $\beta=10$ $\mathrm{Ha}=8$ for NL DNS (top left), QL DNS ($\Lambda=0$) (top right), and GQL DNS with $\Lambda=2$ (bottom left), $\Lambda=3$ (bottom right). Recall that here we choose $r_1$ to be the point halfway between the two cylinders.}
\label{scweak}
\end{figure}

Moreover, even $\Lambda=0$ reproduces axial structure the second cumulants fairly well at $\mathrm{Re}=10000$ (figure \ref{scweak}). The axial wavelength and magnitudes are marginally inaccurate and the radial correlation is focussed at the mid-point instead of across the full gap width, however they are surprisingly good for such a crude approximation to the non-linearity. Interestingly, the correlation structure is still finer than that of the NL DNS, due to the peak at $k\approx20$ that is present in all QL solutions. This structure becomes finer as $\Lambda$ is increased, though the magnitudes of the spurious correlations decrease. 
However, we comment that $\Lambda=0$ does not give universally adequate approximations even in this weakly turbulent case. If we examine $\mathrm{Re}=5000$ (not shown),  the cumulant structure is much less accurate; it is not until $\Lambda=2$, that the axial cumulant structure resembles the NL DNS case. So, whilst for much of this weakly nonlinear parameter regime the standard QL approximation produced adequate  first  and second cumulants, this can not be guaranteed for all $\mathrm{Re}$ -- the $\Lambda=2$ GQL is required to further guarantee this level of accuracy.

\subsubsection{Strong turbulence: $\beta=5$ $\mathrm{Ha}=16$}

\begin{figure}
 \centering
 \includegraphics[width=\textwidth]{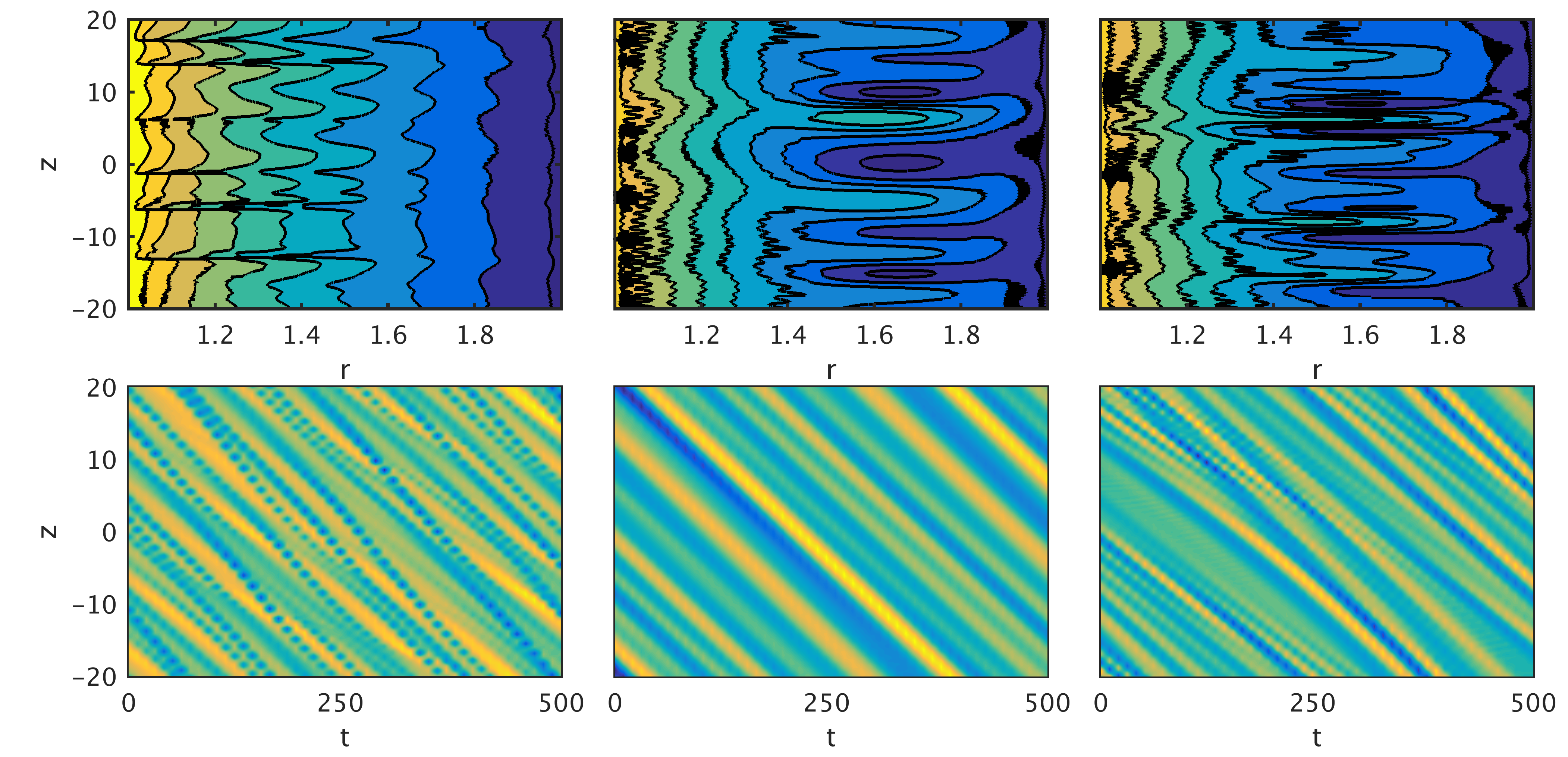}
 \caption{Snapshot contour plots of $u_\phi$ (top) and Hovm\"oller plots (bottom) for $\beta=5$, $\mathrm{Ha}=16$, $\mathrm{Re}=10^4$. Shown on the left is the NL DNS, in the middle is $\Lambda=0$, and on the right is $\Lambda=2$.}
\label{snaphovstrong}
\end{figure}

We next examine the GQL approximation at the set of parameters $\beta=5$, $\mathrm{Ha}=16$, for which we plot the same diagnostic quantities as for NL DNS. From figure~\ref{snaphovstrong}, it is clear that the azimuthal velocity profile is quite different in the $\Lambda=0$ case to that for NL DNS, with the QL approximation failing to reproduce dislocations. Importantly however, we note that the Hovm\"oller plots show that the travelling wave property is preserved, even by the standard $\Lambda=0$ QL approximation. The wavelength, though difficult to measure due to dislocations, can be seen to be within an acceptable range of the NL DNS. Note that if the $\Lambda=2$ GQL approximation is taken, even the nonlinear dislocations are reproduced; a significant improvement over $\Lambda=0$.

\begin{figure}
 \centering
 \includegraphics[width=\textwidth]{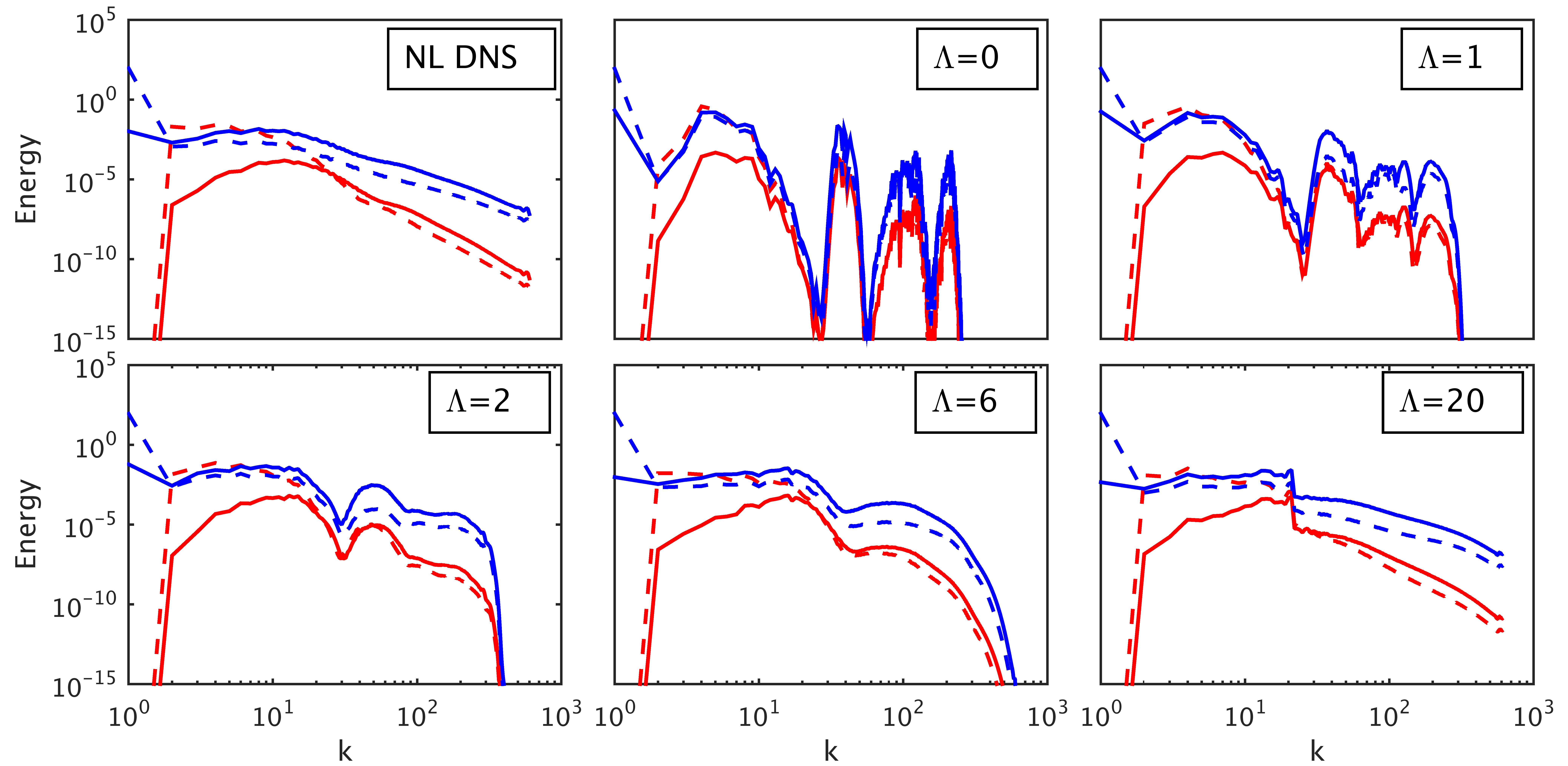}
 \caption{Vertical energy spectra for the various mode truncations at $\beta=5$, $\mathrm{Ha}=16$, $\mathrm{Re}=10000$ with magnetic field (red) and velocity (blue) and with poloidal components (solid) and toroidal components (dashed).}
\label{specstrong}
\end{figure}

Additionally, we compare the energy spectrum to that of the NL DNS (figure \ref{specstrong}). It should be noted that agreement in the lower modes of the spectrum is the most important for the evaluation of the approximation. We observe that the energy in the GQL approximation is located primarily in the lower modes, which peak higher than in the original NL DNS case. Because energy is not being transferred by nonlinear processes the higher modes contain much less, if any, energy and the turbulent cascade is no longer present. 

The most striking features in these plots are the strong peaks in the spectra for GQL at $\Lambda=0$ and $\Lambda=1$, which correspond to a linear instability and its harmonics.  The first peak occurs at a mode number $k=40$,  and the other peaks appear to be the resonance of this at $k=80$ and $k=160$. A key question one may ask is whether this is due to the initial condition -- recall that we started with a fully turbulent state from which the energy could have decayed away in the upper spectra. As noted earlier,  even if the simulations are started from a small perturbation about the basic state these peaks still exist in the spectra, therefore the energy must be transferred outwards to them. For $\Lambda=1$ we still detect these linear peaks, though due to the increased energy scattering off the low modes  some energy has now transferred outwards to the modes between them.

If we turn our attention to $\Lambda=2$, it can be seen that the energy in higher modes becomes more like that in the NL DNS, even if it does decay at a much greater rate for $k>200$. The linear instability peak is still present, however it is but a small protrusion above the energy stored in adjacent modes, and for the modes immediately higher than $k=20$ the spectrum begins to resemble the power law from the NL DNS. The improvement in reproduction of the DNS spectrum over $\Lambda=0$ is clear, yet it can also be seen that increasing $\Lambda>2$ does not yield similarly great increases in accuracy ($\Lambda=6$ produces an almost equivalent spectrum). 
Interestingly there is a sharp decrease in energy around the mode $k=20$, which persists even for $\Lambda=20$, where the rest of the spectra resembles the NL DNS. For this large $\Lambda$, we see that the energy is higher than the NL DNS in the modes slightly lower, and lower in the slightly higher modes, whilst virtually identical to the NL DNS otherwise. 

If we turn our attention to the first cumulants $\langle v \rangle$ in  figure~\ref{fcstrong}, we see that for $\Lambda=0$ (QL) they deviate approximately $10\%$ from the NL DNS solution. As $\Lambda$ is increased the cumulants then approach the profile of NL DNS, with a large improvement in agreement shown between $\Lambda=1$ and $\Lambda=2$, in agreement with what is seen in the energy spectra. The deviation gets smaller as Re is decreased (not shown) but decreasing Re does not lead to a better approximation; the lower Reynolds number simply means that the cylinder is rotating with lower angular velocity and thus the overall perturbation in the velocity profile is smaller, with the percentage error for QL staying approximately the same.

We also compare plots of the second cumulant, $\langle v' v' \rangle$. Recall we choose to display the cross-correlation with the mid-point of the gap between the cylinders. The choice of Re has a slight influence on the second cumulants though does not significantly alter their axial structure. Thus, we choose to examine the case when $\mathrm{Re}=7000$ to illustrate the behaviour of the second cumulants in the GQL approximation.

\begin{figure}
 \centering
 \includegraphics[width=0.7\textwidth]{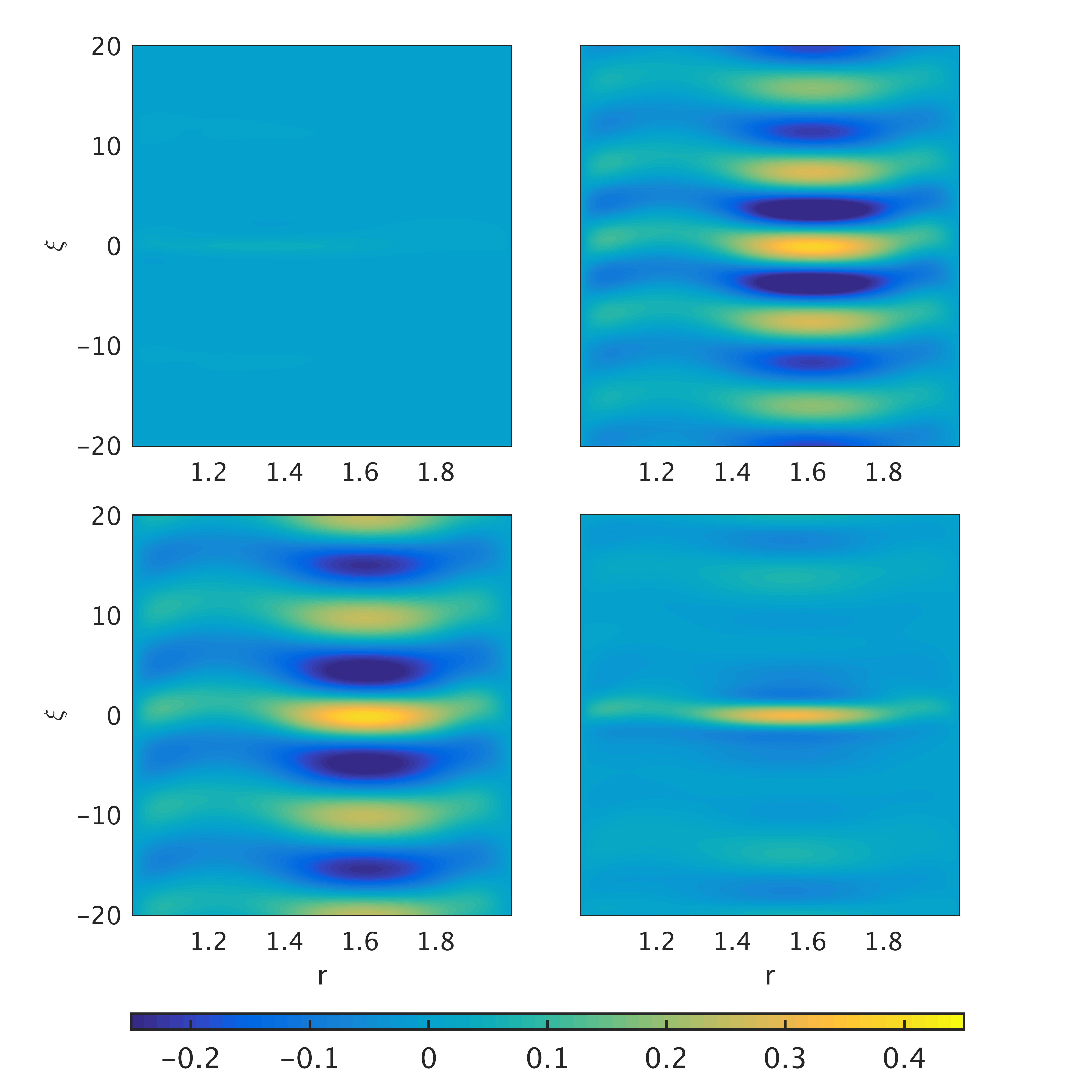}
 \caption{Second cumulants $\langle v' v' \rangle $, as in Figure 4, at a number of mode truncations for the parameter regime $\beta=5$ $\mathrm{Ha}=16$, $\mathrm{Re}=7000$. Shown are; NL DNS (top left), QL DNS ($\Lambda=0$) (top right), and GQL DNS with $\Lambda=1$ (bottom left), $\Lambda=2$ (bottom right).}
 \label{scstrongvv}
\end{figure}
It is immediately clear from Figure~\ref{scstrongvv} that the second cumulants for $\Lambda=0$ and $\Lambda=1$ are inadequate reproductions of the NL DNS case --- for these approximations there is an uncharacteristically large observed axial correlation. Indeed, if we take $\Lambda=0$ we see that the correlation has a wavenumber of 5, whereas the NL DNS is strongly correlated with one axial location, and is very weakly (or un-) correlated with the rest of the domain. Interestingly, $\Lambda=1$ has correlation of wavenumber 4, possibly associated with the presence of waves travelling axially (see also \citet{mct16}). Recall that the localisation of the second cumulant appeared to be associated with the absence of a strong spectral peak. As such, by $\Lambda=2$ the axial localisation of the second cumulant has been re-established.

\begin{figure}
 \centering
 \includegraphics[width=0.7\textwidth]{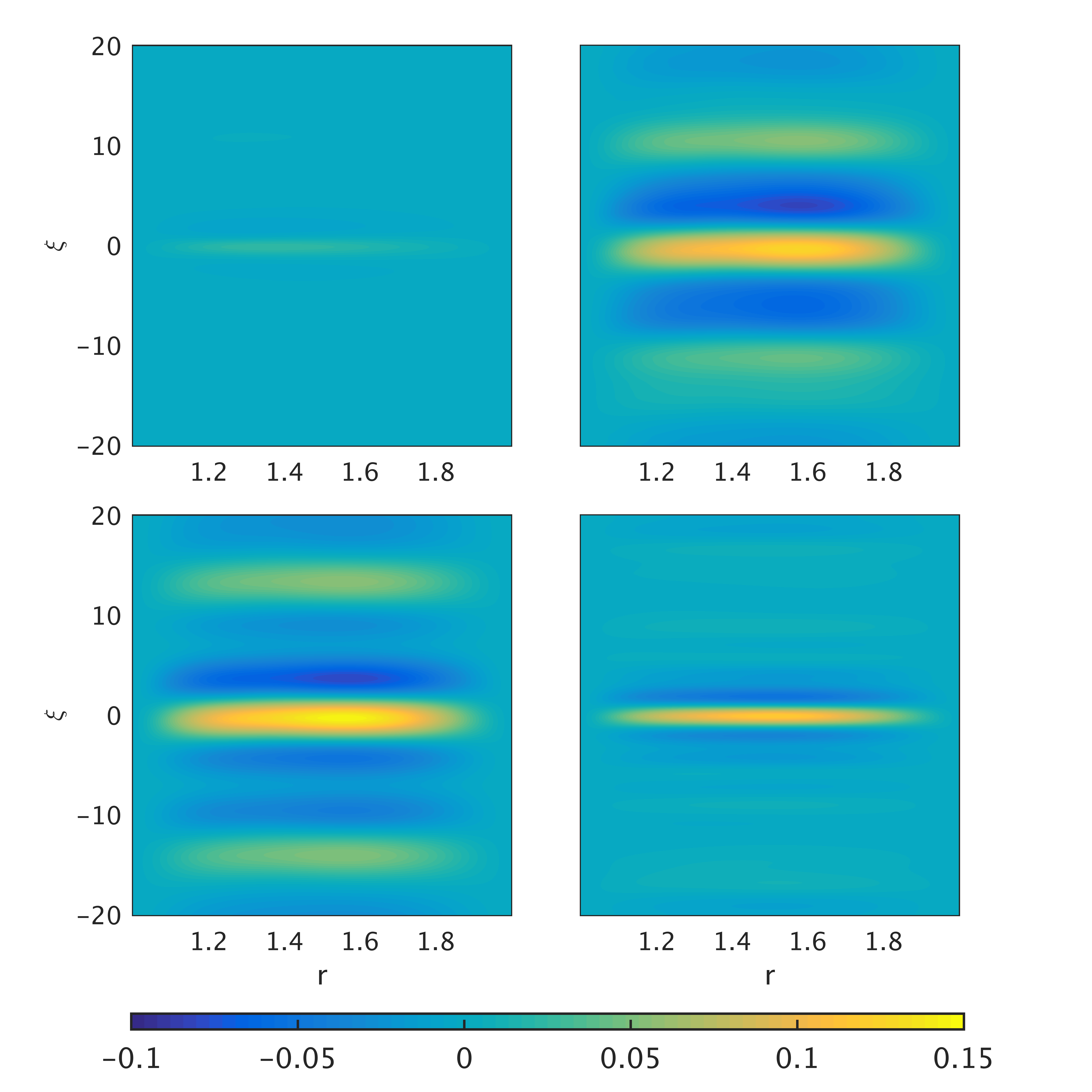}
 \caption{As for Figure~\ref{scstrongvv} but for cross cumulants $ \langle v' \psi' \rangle $. }
\label{scstrongvpsi}
\end{figure}

We note similar structure in the cross cumulants $\langle v'\psi' \rangle$, which are plotted in figure~\ref{scstrongvpsi}. A similar overcorrelation is seen for $\Lambda=0$ and $\Lambda=1$ , though $v'$ is out of phase with $\psi'$. At $\Lambda=2$ we see that the structure of the NL DNS is regained, with all but the axial midpoint being very weakly correlated. Note how now the radial correlation is much more consistent across the NL DNS and GQL approximations than previous.

\subsubsection{$\beta=2.5$ $\mathrm{Ha}=16$: moderate turbulence}

As mentioned previously, this parameter set is much less turbulent than $\beta=5$, $\mathrm{Ha}=16$, with fewer dislocations in the NL DNS. Thus, it can inform us of the effect that $\mathrm{Re}_c$ has on the viability of the GQL, in a similar fashion to the reduced Re cases previously. Much like before, it can be found that the fundamental travelling wave property is reproduced even for $\Lambda=0$, and it is not until $\Lambda =2$ that nonlinear nonlinear dislocations are again found in Hovm\"oller plots.

If we examine the first cumulants in figure~\ref{fcmt} we see that they have profiles that are very close to that of the NL DNS, whatever the degree of approximation --- though improvement is seen as $\Lambda$ is increased. We briefly comment that the energy spectra and second cumulants show similar behaviour to those for the strong turbulence case described above, with GQL beginning to yield a good approximation by $\Lambda=5$. Hence we conclude that even if QL appears to give an adequate approximation on examination of the mean flows, GQL may be required for accurate representation of the fluctuations.

\section{Conclusion}

In this paper we have extensively examined the effectiveness of the Generalised Quasilinear Approximation (GQL) for the case of an axisymmetric HMRI for a number of different parameter regimes, with flows varying from weakly  to strongly turbulent. Establishing the approximation's effectiveness is important not only for determining which interactions play a key role in the dynamics of the system, but also for determining the minimum ingredient of a statistical theory (such as Direct Statistical Simulation) that is necessary to reproduce the low-order statistics of the system.

For weakly turbulent systems we find no fundamental problems with the possibility of DSS and the HMRI -- even for the QL case of $\Lambda=0$ the travelling wave behaviour is reproduced, even if some of the other properties are of varying degrees of accuracy.  However  when the system is more turbulent we have found that for the GQL approximation at  $\Lambda=2$, much more accurate results can be obtained, and that increasing $\Lambda$ further yields diminishing returns, with a much lesser improvement in accuracy per extra mode included. We have seen that $\Lambda=2$ gives a much improved spectra compared with $\Lambda=0$. The various cumulants further reinforce this; $\Lambda=0$ for the most part yields over-correlated second cumulants, and it is not until $\Lambda=2$ that the correlation structures properly resemble the NL DNS.

The choice of parameter regime has a knock--on effect in the accuracy of the GQL approximation, with the degree of supercriticality an important factor. However, for the majority of the parameter space the influence that $\beta$, Ha, and Re have on the approximations has little effect on the relative accuracy of the different $\Lambda$ approximations; in each regime it was found that increasing to $\Lambda=2$ offers a marked improvement over the $\Lambda=0$ case whilst increasing $\Lambda>2$ yields much smaller increases in accuracy. Special points, such as the weakly nonlinear $\beta=10$, $\mathrm{Ha}=8$ may be approximated to a good degree of accuracy using even the $\Lambda=0$ approximation. However, these are but a small section of the total parameter space.

This leads us to two main conclusions. Firstly, as the HMRI is the subject of much current research, it may be fruitful to perform direct statistical simulations of the instability. Having shown that using a GQL approximation may yield accurate results, the equivalent cumulant expansion to the $\Lambda=2$ case may be used to give further insight into the instability.

Secondly, and most importantly, this work shows that for some problems it may be advantageous to use cumulant expansions other than the quasilinear CE2 expansion to conduct direct statistical simulation. As mentioned in the introduction, previous research has shown that DSS via the CE2 expansion (equivalent to $\Lambda=0$), whilst accurate for problems such as out of equilibria jets, can also prove inaccurate. We have shown that for the HMRI, where taking a standard QL approximation would have been inadequate, even the addition of two more large scale modes in the $\Lambda=2$ approximation yields markedly more accurate solutions. It is logical to conclude that a similar $\Lambda\ge2$ cumulant expansion would allow for accurate DSS in other flows, without compromising the computational advantage of DSS over DNS. These results reinforce those of \citet{mct16} who found that the GQL approximation may yield significant enhancement over QL for the hydrodynamic problem of jet formation on a spherical surface or $\beta$-plane. 

Finally, we note that while we have here only considered the axisymmetric HMRI, for only slightly larger Hartmann numbers there is also a non-axisymmetric analog, the so-called azimuthal MRI (AMRI). This mode undergoes a similar switch in scaling from $\mathrm{Rm}$ to $\mathrm{Re}$ as in the SMRI$\to$HMRI transition \citep{hollerbach2010}, and has by now also been observed in the PROMISE experiment \citep{seilmayer}. 3D numerical modelling of these modes is only just beginning \citep{guseva}, but a GQL analysis as presented here would obviously also be of considerable interest, with the separation into different scales then possible in both axial and azimuthal directions. Indeed encouragement for this enterprise is obtained from results for the hydrodynamic rotating Couette problem which indicate that GQL remains an attractive approximation when averaging over two spatial directions (in that case streamwise and spanwise directions).

\section*{Acknowledgements}

AC was supported by an STFC studentship, RH and ST were supported by STFC Grant No. ST/K000853/1, and BM was supported by NSF Grant No. DMR-1306806.

\end{document}